\documentclass[10pt,english]{article}
\usepackage[LGR,T1]{fontenc}
\usepackage[utf8]{inputenc} 

\usepackage{xcolor}
\usepackage{geometry}
\usepackage[unicode=true,bookmarks=true,bookmarksopen=true,breaklinks=true,pdfborder={0 0 0},pdfborderstyle={},colorlinks=true,allcolors=blue]{hyperref}

\usepackage{setspace}
\usepackage[natbib, style=numeric, uniquelist=false, uniquename=false, sortcites=true]{biblatex}
\AtEveryBibitem{\clearfield{month}}
\AtEveryCitekey{\clearfield{month}}
\AtEveryBibitem{\clearfield{labelmonth}}
\AtEveryCitekey{\clearfield{labelmonth}}
\AtEveryBibitem{\clearfield{day}}
\AtEveryCitekey{\clearfield{day}}
\AtEveryBibitem{\clearfield{labelday}}
\AtEveryCitekey{\clearfield{labelday}}

\usepackage{nth}
\usepackage{comment}
\usepackage{color}
\usepackage{float}
\usepackage[position=top]{subfig}
\usepackage{amsmath}
\usepackage{amsthm}
\let\amsnewtheorem\newtheorem %
\usepackage{amsfonts}
\usepackage{mathrsfs}  %
\usepackage{amssymb}
\usepackage{stmaryrd}
\usepackage{graphicx}
\usepackage{appendix}
\usepackage[normalem]{ulem}   %
\usepackage{bookmark}
\usepackage{sepfootnotes}
\usepackage{bm}
\usepackage{enumitem}

\usepackage[ruled, algosection]{algorithm2e}
\usepackage{booktabs} %

\let\originalleft\left
\let\originalright\right
\renewcommand{\left}{\mathopen{}\mathclose\bgroup\originalleft}
\renewcommand{\right}{\aftergroup\egroup\originalright}

\usepackage{mathtools} 
\DeclarePairedDelimiter\abs{\lvert}{\rvert}%
\DeclarePairedDelimiter\norm{\lVert}{\rVert}%
\makeatletter
\let\oldabs\abs
\def\abs{\@ifstar{\oldabs}{\oldabs*}}
\let\oldnorm\norm
\def\norm{\@ifstar{\oldnorm}{\oldnorm*}}
\makeatother

\newcommand{\E}{\mathbb E}

\usepackage{bbm}

\usepackage{multirow, colortbl}
\usepackage{algorithmic}
\usepackage{xpatch}

\theoremstyle{definition}

\theoremstyle{remark}

\amsnewtheorem*{excont}{Example \continuation} %
\newcommand{\continuation}{??}

\makeatletter
\let\oldsection\section
\renewcommand{\section}{\@ifstar{\sectionstar}{\oldsection}}
\NewDocumentCommand{\sectionstar}{s o m}{%
	\oldsection*{#3}%
	\addcontentsline{toc}{section}{#3}%
}

\let\oldsubsection\subsection
\renewcommand{\subsection}{\@ifstar{\subsectionstar}{\oldsubsection}}
\NewDocumentCommand{\subsectionstar}{s o m}{%
	\oldsubsection*{#3}%
	\addcontentsline{toc}{subsection}{#3} %
}
\makeatother

\title{Why Data Anonymization Has Not Taken Off}

\author{Matthew J. Schneider\footnote{Drexel University \& Boston College, mjs624@drexel.edu}, James Bailie\footnote{Harvard University, jamesbailie@g.harvard.edu} \ and Dawn Iacobucci\footnote{Vanderbilt University, dawn.iacobucci@vanderbilt.edu}}

\date{\today}

\begin{document}

\maketitle

\begin{abstract}
Companies are looking to data anonymization research -- including differential private and synthetic data methods -- for simple and straightforward compliance solutions. But data anonymization has not taken off in practice because it is anything but simple to implement. For one, it requires making complex choices which are case dependent, such as the domain of the dataset to anonymize; the units to protect; the scope where the data protection should extend to; and the standard of protection. Each variation of these choices changes the very meaning, as well as the practical implications, of differential privacy (or of any other measure of data anonymization). Yet differential privacy is frequently being branded as the same privacy guarantee regardless of variations in these choices. Some data anonymization methods can be effective, but only when the insights required are much larger than the unit of protection. Given that businesses care about profitability, any solution must preserve the patterns between a firm’s data and that profitability. As a result, data anonymization solutions usually need to be bespoke and case-specific, which reduces their scalability. Companies should not expect easy wins, but rather recognize that anonymization is just one approach to data privacy with its own particular advantages and drawbacks, while %
the best strategies jointly leverage the full range of approaches to data privacy and security in combination.%

\end{abstract}

\subsection*{Data Security vs. Data Anonymization}
Organizations have well-established 
security tools for collecting, storing, and processing personal data. They are standard and readily scalable across industries, regardless of the type of data being used. %
Firewalls, access controls, automated auditing, and intrusion detection can be employed to prevent malicious activity; data can be stored and transmitted in an encrypted form so that it is impossible to read without proper credentials; and multi-factor authentication with strong passwords can stop unauthorized access. These technologies are uncontroversial and straightforward to deploy out of the box. %
There are few barriers to their adoption -- an organization simply needs to choose one of the many providers on the market, and they will quickly have a secure %
IT system complying with the best security practices. %

However, good data privacy requires more than good security. %
The Facebook–Cambridge Analytica scandal, for example, did not arise from a security breach. Rather, it was the result of %
unethically collecting and sharing personal information---the data of millions of Facebook users were harvested without their consent and %
subsequently disseminated to third parties with the understanding that these data would be used to influence democratic elections \citep{rosenbergTrumpConsultants2018,hernCambridgeAnalytica2019}. Facebook and Cambridge Analytica were not hacked; indeed the data were collected and shared through secure channels. Yet, the privacy and trust of millions of people were nonetheless violated. Many such breaches are occurring across a diverse range of industries, including healthcare, transport and entertainment \citep{narayananRobustDeanonymizationLarge2008, culnaneStopOpenData2019, sweeneySimpleDemographicsOften2000}. 

To comply with privacy regulation and maintain a positive public image, organizations must therefore ensure that the data they acquire is appropriate and justified, and that any sharing of data is done responsibly. %
\emph{Data anonymization} -- the process of making personal data \emph{less personal} by adding noise, masking, or removing variables -- is one tool that help companies meet these goals \citep{elliotAnonymisationDecisionMakingFramework2020, gadottiAnonymizationImperfectScience2024}. 
By applying anonymization techniques when collecting or disseminating data, 
organizations can limit the exposure of sensitive personal information. This can reduce privacy risks, support regulatory compliance, and generally contribute to the responsible handling of personal data.

Since IT security tools such as encryption largely operate automatically in the background of today's digital systems, organizations might anticipate a similar level of convenience with data anonymization methods. 
It is tempting for a business to assume that it can meet its ethical and regulatory obligations by purchasing off-the-shelf anonymization software and plugging it into its existing data pipelines.
The expectation is that anonymization can be applied as seamlessly and painlessly as data security, with turnkey solutions that 
guarantee privacy compliance and ethical data collection. %
As a result, many business executives, policy makers, and legislators believe simple, risk-free anonymization is possible \citep{sweeneySayingItsAnonymous2018}.
Yet the situation could not be further from the truth. In this article, we explain intuitively and theoretically why data anonymization is costly, requires making a complex set of choices, and does not transfer well across industries or use cases. We show that anonymization is on a spectrum, cannot reduce the risk of re-identification/de-anonymization to zero, and requires context-specific decisions on how data will be used and what must be protected.

\subsection*{Overview of Data Anonymization}

The collection of personal data is ubiquitous in modern society. Companies have access to an unprecedented amount of sensitive, private information, along with personally identifiable information that can be used to relate sensitive data to its subject. Examples of personally identifying information include direct or unique identifiers -- such as names and social security numbers -- as well as what are called `quasi-identifiers,' variables such as date of birth, ZIP code, and gender, which can be combined to identify individuals.

One very basic approach to data anonymization is to delete personally identifying information from a dataset before it is shared with third parties. %
However, it is now widely recognized that %
this approach frequently does not provide sufficient protection on its own \citep{sweeneySimpleDemographicsOften2000, barbaroFaceExposedAOL2006, narayananRobustDeanonymizationLarge2008, homerResolvingIndividualsContributing2008, ohmBrokenPromisesPrivacy2010, demontjoyeUniqueShoppingMall2015, dworkExposedSurveyAttacks2017}. Variables that may appear innocuous often turn out to reveal sensitive information. Data that may not appear to be identifiable can be linked to other datasets. %
As such, it can be difficult to determine what pieces of information count as quasi-identifiers and need to be removed. It may even be possible to argue that, in the right context, any one single variable in a dataset could be used to identify data subjects -- in which case no data could be released at all.

Because what constitutes personally identifying information is ambiguous, there has been an increasing recognition of the need for more robust anonymization methods.
Today, data anonymization comprises a broad class of privacy enhancing technologies (PETs) and  variety of ideas, including data swapping \citep{daleniusDataswappingTechniqueDisclosure1982}, input and output perturbation \citep{thompsonMethodologyAutomaticConfidentialisation2013, shlomoStatisticalDisclosureControl2007, okeefeSummaryAttackMethods2013}, synthetic data \citep{drechslerSyntheticDatasetsStatistical2011}, cell suppression \citep{coxSuppressionMethodologyStatistical1980}, differential privacy \citep{dworkCalibratingNoiseSensitivity2006}, formal privacy \citep{panelonapproachestosharingblendeddataina21stcenturydatainfrastructure21stCenturyNational2024}, aggregation \citep{duncanDisclosurelimitedDataDissemination1986}, $k$-anonymity \citep{sweeneyKAnonymityModelProtecting2002}, the $p$\% rule \citep{coxLinearSensitivityMeasures1981}, randomized response \citep{warnerRandomizedResponseSurvey1965} and statistical disclosure control \citep{hundepoolStatisticalDisclosureControl2012}.
Despite their variety, all data anonymization methods share a unifying goal: they are technical tools designed to \emph{reduce} the amount of personal information contained in the data. However, they do not eliminate all personal information in the data;
rather they restrict the granularity or resolution of statistical inferences that can be drawn. In doing so, they aim to preserve the utility of the data for some use cases while constraining the extent that personal information can be learned.

\subsection*{Data Anonymization: A Baskin-Robbins of Choices}

Drawing the boundary between permissible and impermissible inferences poses significant challenges. Organizations must make difficult design choices when implementing data anonymization methods, often balancing protection against potential impacts on business value. For example, consider the case of a retailer anonymizing customer data before sharing it with an external party:

\begin{enumerate}
    \item \textbf{Domain}: the retailer must choose which data set(s) to anonymize and share. For example, the retailer may share a point-of-sale transaction data set with two firms -- a market research firm to improve promotion strategies and a private equity firm to predict hourly foot traffic for nearby portfolio companies. Additional data sets could also be shared, including employee behaviors, energy usage, or supply chain orders.
    \item \textbf{Unit}: the retailer must choose the unit of protection. %
    Should the level of anonymization be at the level of transactions, consumers, SKUs, brands or retail stores?
    \item \textbf{Scope}: the retailer must choose what variables the anonymization protection should extend to. Should the protection extend to SKUs, brands, or product categories? Should the protection also extend to consumer demographics and prices? Note that robust anonymization requires that the influence of a single protection unit on the outputs (e.g., business insights) of these variables is small.
    \item \textbf{Anonymization strength}: the retailer must choose the \textbf{standard} and \textbf{intensity} of protection. These are technical requirements relating to how anonymization is measured (the standard) and how much anonymization is added (the intensity).
\end{enumerate}

These choices are contextually driven, and many variations can emerge -- far more than the 31 flavors originally marketed by Baskin-Robbins. Each variation of \{domain, unit, scope, standard\} requires a bespoke technological solution, and by definition, this reduces scalability across applications and industries. 
    For example, an anonymization solution for a retailer with the first three choices \{domain = weekly store sales data, scope = brand/prices/sales, unit = store ID\} may include a synthetic data solution in order to generate fake sales data \citep{schneiderFlexibleMethodProtecting2018}. 
    By comparison, a pharmacy network with choices \{domain = monthly prescription data, scope = brand/prescription counts, unit = physician\} may include an anonymization solution which aggregates physicians’ prescriptions while maintaining brand insights at the state-level \citep{li2023reidentification}.

With all these choices, data anonymization clearly is not a binary yes/no, but rather there are many different flavors of anonymization, none of which guarantee safety \citep{hartzogAnonymizationDebateShould2017}. 
Applying a data anonymization method -- regardless of the method -- does not mean that your data is now `anonymized'; it only means that the amount of personal information contained in the data has been reduced. These decisions involve tradeoffs -- stricter settings reduce the risk of exposing personal information, but all data sharing carries some risk, no matter how thoroughly anonymized.

As companies are always looking to maximize their profits, 
any anonymization must preserve patterns between the companies' data and their profitability. Thus, if employed, anonymization solutions -- unlike data security solutions -- are usually case-specific and not readily deployable across other use cases or industries. Accordingly, these customized solutions can be costly and confusing to business professionals -- or even privacy experts themselves!

\subsection*{Theoretical Anonymization: 200+ Flavors of Differential Privacy}

From a theoretical perspective, the most common anonymization approaches that preserve profitability can be considered as ad hoc because they do not have a mathematical guarantee of privacy. This is why academia has downplayed usefulness \citep{blanco-justiciaCriticalReviewUse2022} and almost entirely
focused on a theoretical guarantee of data anonymization -- called \emph{differential privacy} \citep{dworkCalibratingNoiseSensitivity2006} -- a term that has become the `good housekeeping' seal of privacy. Companies such as Apple, LinkedIn and Microsoft commonly advertise that their data are differentially private to position their brands among consumers \citep{greenbergHowOneApples2017, rogersLinkedInAudienceEngagements2021, edmondHowStatisticalNoise2020}. 

Although differential privacy has some good theoretical properties -- for example, composition (i.e., one differentially private data release will not completely undo the protection imbued in another data release) --  
the theory is not always as ironclad as it sounds. It turns out that varying any of the four choices of \{domain, unit, scope, standard\} leads to a different definition of differential privacy, even though they all are mathematically rigorous \citep{bailieRefreshmentStirredNot2024a}. The theoretical development of new differentially private methodologies and their branding in practice has truly become bigger than the `Baskin-Robbins' of data anonymization. By varying the choices of \{domain, unit, scope, standard\}, researchers have created over 200 different flavors of differential privacy (and counting!), with each implying a different meaning of anonymization, and each appropriate only in specific contexts \citep{desfontainesSoKDifferentialPrivacies2020}. 

Yet differential privacy is often being branded as the same ironclad guarantee regardless of the chosen specification -- i.e., regardless of the chosen settings for all five parameters (the domain, unit, scope, standard and intensity of protection)
\citep{dworkDifferentialPrivacyPractice2019, desfontainesListRealworldUses2023}. The problem is that many of these %
specifications are ``privacy mostly in name'' only \citep{dworkDifferentialPrivacyPractice2019}, so one needs to assess the chosen specification on a case-by-case basis to determine if it provides sufficient protection. However, it is difficult (even for privacy experts) to assess which specifications are appropriate for a given data release, and which are not \citep{nanayakkaraWhatAreChances2023a, heffetzWhatWillIt2022}. Worse still, 
some companies are advertising their use of differential privacy without disclosing the specification they are using \citep{desfontainesListRealworldUses2023} -- but
there is no guarantee of any protection (let alone a sufficient level of protection) if the data collector's chosen differential privacy specification is not disclosed. %
Moreover, differential privacy (and data anonymization more generally) is not a panacea for all the privacy harms that arise in data processing \citep{seemanPrivacyUtilityDifferential2023, coleDesigningAccessDifferential2020, seemanDifferentialPrivacyPublic2025}. As such, companies still need to recognize and mitigate the risk of these harms even if they use gold-standard differential privacy protections.

\subsection*{Barriers to Adoption}

Firms want straightforward and easy to implement compliance solutions. %
Data anonymization has emerged as a potential solution which seems simple (typically it involves adding some noise to the aggregate statistics you want to release). However, it is anything but simple in practice.

Differential privacy, despite its theoretical rigor, exemplifies anonymization's implementation difficulties.
In addition to choosing an appropriate flavor, deploying differential privacy requires a deep understanding of how the data was generated \citep{huProvablePrivacyNonprivate2024a, drechslerComplexitiesDifferentialPrivacy2024}, and most existing DP implementations have required new, bespoke methodologies developed by leading researchers in the field \citep{desfontainesListRealworldUses2023}. Adopting differential privacy as a compliance solution is further complicated by the Baskin-Robbins of incompatible flavors (e.g., exchanging data with business partners can be difficult when one firm relies on one combination of \{domain, unit, scope, standard\} and the other firm relies on another).

After two decades of research, large-scale deployments of differential privacy remain limited to narrow domains and well-funded institutions with the financial resources to support large-scale research and development -- the 2020 US Census being perhaps the most prominent example, where adopting differential privacy required a large team of experts more than five years to implement for a nationwide seven-question dataset and ``degraded the value of the \ldots\ data products in terms of timeliness and quality'' \citep{nationalacademiesofsciencesAssessing2020Census2023}. The absence of established norms for selecting from the Baskin-Robbins of differential privacy flavors forces practitioners to navigate complex tradeoffs without clear guidance \citep{coleDesigningAccessDifferential2020}, often resulting in configurations that serve as ``privacy theater'' rather than meaningful protection \citep{smartUnderstandingRisksPrivacy2022}. This has raised concerns among regulators about the potential for privacy-washing, with bodies like the UK Information Commissioner's Office expressing caution about differential privacy's use for regulatory compliance \citep{ukinformationcommissionersofficeUKGDPRGuidance2023}.   

There are some on-going initiatives to combat some of the challenges, but there is much progress which still needs to be made \citep{cummingsAdvancingDifferentialPrivacy2024, cummingsCenteringPolicyPractice2023, heffetzWhatWillIt2022}, and it is unclear how successful they will ultimately be since some of the adoption difficulties are fundamentally intractable. 
Firstly, while software packages like OpenDP \citep{Opendp_white_paper_11may2020pdf} attempt to democratize access to differentially private algorithms, their sustainability and scope remain uncertain -- OpenDP itself has yet to achieve a stable release after five years of development and funding by multiple large institutions, 
highlighting the persistent gap between academic innovation and practical implementation. Besides, deploying data anonymization requires a lot more than code \citep{cummingsCenteringPolicyPractice2023, seemanPrivacyUtilityDifferential2023, drechslerDifferentialPrivacyGovernment2023}; most of the work cannot be outsourced into a software package. Secondly, although the research community has started to investigate how to decide what differential privacy specifications are appropriate for a given data release, they only address how to set the fifth parameter of a differential privacy specification (the intensity of protection), while ignoring how to choose the first four (the domain, unit, scope and standard) \citep{kazanPrioritizingPrivacyBayesian2025, abowdEconomicAnalysisPrivacy2019, leeHowMuchEnough2011, hsuDifferentialPrivacyEconomic2014}. Since the meaning of the fifth parameter depends on the choices for the first four, the real problem -- setting all five interrelated parameters simultaneously -- is much more complex than what is currently being tackled by the academic literature.

The adoption of data anonymization is hard and often driven by brand positioning \citep{steed2024adoption}.  Implementing anonymization in practice involves costly decisions that extend far beyond selecting algorithms. Organizations must choose whether to apply anonymization during data collection, within the analytical pipeline, or at the point of sharing \citep{drechslerComplexitiesDifferentialPrivacy2024}. For example, a untrusted tech company may apply anonymization locally on customers' phones so they never see the real data, while a reliable institution such as the US Census Bureau may only apply anonymization immediately before disseminating their data. Each approach presenting distinct tradeoffs between statistical efficiency and privacy protection. The fundamental challenge lies in modifying existing data pipelines \citep{bailieWhoseDataIt2024}, ensuring downstream users can still extract meaningful insights and use the data as they envisioned.  Scaling from academic toy examples to complex real-world datasets that require bespoke solutions and significant technical expertise reduces adoption of data anonymization across companies. 

Emerging solutions -- such as synthetic data generation from generative adversarial networks \citep{anandUsingDeepLearning2023, ponteWheresWaldoFramework2024} and large language models -- offers potential pathways to greater scalability by automating aspects of the anonymization process. However, they have their own limitations. For one, synthetic data cannot simply replace real data in existing analytical workflows without modifying the profitability of data-driven models. In fact, whenever an anonymization method is implemented, downstream uses of the data will need to be modified to account for the anonymization \citep{gongTransparentPrivacyPrincipled2022}. Furthermore, these AI solutions inherit privacy risks from their training data \citep{tramerPositionConsiderationsDifferentially2024}. Despite promises from some \citep{stadlerSyntheticDataAnonymisation2022}, it is not a given that synthetic data necessarily provides any protection \citep{nasrScalableExtractionTraining2023, jordonSyntheticDataWhat2024}. Rather, for each synthetic dataset, one needs to demonstrate before releasing it that it indeed masks the confidential data it was generated from -- a tasks which is not easy for sophisticated machine learning algorithms such as generative adversarial networks and large language models \citep{carliniSecretSharerEvaluating2019, jordonSyntheticDataWhat2024}. Finally, evidence suggests that organizational adoption of anonymization techniques is often driven more by branding considerations than genuine privacy protection commitments \citep{steed2024adoption}, indicating that even when technical solutions mature, the broader ecosystem of incentives and expertise required for responsible implementation remains underdeveloped.
 
\subsection*{Where Do We Go From Here?}

There is a belief in some sectors that data can retain its utility at the highest levels of anonymization. Yet this is simply not true. There is a fundamental trade-off between the level of anonymization and the utility of the data. Given the difficulty balancing anonymity and profitability in practice, most companies have been downplaying anonymization solutions despite the widespread academic interest in data anonymization. Rather, companies continue to use real consumer-level data within secure cloud environments (such as AWS or Azure), or cloud-based AI and data platforms to generate insights (such as Databricks, Snowflake, or Palantir Foundry). 

The truth is that data anonymization (including differential privacy) may be effective in only a small subset of commercial use cases -- namely, scenarios where the insights required (e.g., state-wide statistics) are much broader than the unit of protection (e.g., consumers) \citep{kiferNoFreeLunch2011}. For example, anonymization solutions for four-decimal-place GPS coordinates are probably not going to work well (the resolution of analysis is too high), but anonymization solutions for reporting disease prevalence in major cities across large population subgroups (a low resolution) can be effective. Relative to the size of the data, more noise must be added to protect a company with data on 10 consumers compared to a company with 10 million consumers for the same level of protection. This adversely affects smaller companies and under-represented groups, destroying their data's usefulness \citep{santos-lozadaHowDifferentialPrivacy2020, nationalacademiesofsciencesIdentificationRuralSpecial2020, ArchivencaiorgPolicyresearchcenterResearchdata}. 

It will be hard for companies to accept expensive, bespoke methods that destroy the original reason the consumer data was collected in the first place, but this may be their only option if they would like to apply data anonymization. Since consumer-level insights and consumer-level anonymization are likely not possible at the same time, companies, researchers and regulators should not over-promise nor over-rely on data anonymization. There are other, simpler data privacy and security solutions -- such as aggregation, encryption, retention and deletion policies, vetting users, notice and consent processes, enclaves and secure access systems, query auditing and data user agreements -- which are often more palatable, deployable and scalable while still complying with privacy regulation. 

On their own, these approaches may not be sufficient -- unfortunately, there are no silver bullets here. But nevertheless they may each provide some benefit, so that they can be combined into a satisfactory solution. Therefore we recommend that firms, academics and policy-makers take a comprehensive approach to data privacy. Holistic models for data sharing -- such as Contextual Integrity and the Five Safes \citep{nissenbaumContextualIntegrityData2019, nissenbaumPrivacyContextTechnology2010, greenPresentFutureFive2023, desaiFiveSafesDesigning2016} -- are effective guides for navigating data privacy quagmires. Rather than perpetuating the myth of the binary between personal and anonymized data, policy should follow a risk- and process-based approach, as has proven effective for IT security \citep{rubinsteinAnonymizationRisk2016, hartzogAnonymizationDebateShould2017}. But first and foremost, the community needs to acknowledges the real limitations of anonymization techniques, the unattainable goal of perfect, risk-free anonymization, the Baskin-Robbins of anonymization flavors, the challenges in their adoption and their applicability only to a limited set of privacy harms.

\subsection*{Acknowledgments}
The authors thank the anonymous reviewer for their helpful feedback.

\vspace{0.5em}
\noindent Data availability statement: There was no data used in this manuscript.

\noindent Ethics and consent to publish declarations: Not applicable.

\noindent There are no competing interests or financial conflicts of interest.

\noindent Funding declaration: JB gratefully acknowledges partial financial support from the Australian-American Fulbright Commission and the Kinghorn Foundation.

\begingroup
\sloppy
\printbibliography
\endgroup

\end{document}